# Multihop Adjustment for the Number of Nodes in Contention-Based MAC Protocols for Wireless Ad hoc Networks

Ash Mohammad Abbas

Department of Computer Engineering
Aligarh Muslim University
Aligarh - 202002, India
`am.abbas.ce@amu.ac.in`

**Abstract.** The number of contending neighbors of a node in a multihop ad hoc network has to be adjusted while analyzing the performance of the network such as computing the end-to-end delays along a path from a given source to a destination. In this paper, we describe a method to adjust the number of contending neighbors of a node in a multihop wireless ad hoc network. Our method is based on the minimum number of neighbors that has to be common between two consecutive nodes along a path. We derive an analytical expression for the adjustment factor.

## 1 Introduction

The IEEE 802.11 DCF (Distributed Coordination Function) and IEEE 802.11e EDCA (Enhanced Distributed Channel Access) have potentials to be used as *medium access control* (MAC) layer protocols in wireless ad hoc networks. IEEE 802.11 DCF and IEEE 802.11e EDCA are based on *Carrier Sense Multiple Access with Collision Avoidance* (CSMA/CA), which is a contention based MAC protocol. In other words, a station that has frames to transmit senses the carrier and if it finds the carrier free for a certain time duration, it starts transmitting the frame. After finishing the transmission it waits for an acknowledgment from the receiver. If another station starts transmitting when a tagged station is transmitting, the collision is said to occur. If it does not receive the acknowledgement, it tries to transmit the frame again after going through the backoff procedure, until a retry limit is reached. When the number retransmissions exceeds the retry limit, the frame is discarded.

Many researchers have proposed analytical models for evaluating the performance of IEEE 802.11 DCF and IEEE 802.11e EDCA. For example, the performance of IEEE 802.11 DCF has been evaluated in [2], [3], [4]. Models for evaluating the performance of IEEE 802.11e EDCA are proposed in [5], [6], [7], [8]. All these studies consider a single cell wireless local area network (WLAN) consisting of, say $n$, nodes. Since the network does not have multiple hops, therefore, all these $n$ nodes contend for the channel. A common assumption among all these models is that collision probability and the probability of transmission (or transmission attempt rate [3]) is a function of the number of nodes contending for the channel. The analysis of the average delay and average throughput is based on the above assumption. One may ask a question: What should one do in a multihop network (such as an ad hoc network)? Should one multiply the average delay by the number of hops to get the average end-to-end delay? The answer of the question is not so trivial and straight forward. To answer it, one has to consider a model for the ad hoc network.

In case of an ad hoc network, where there are often multiple hops between a given source and a destination, the situation is different. Only nodes that are lying within the carrier sense range of a node contend for the channel. Specifically, the number of nodes contending for the channel should not be counted more than once if the respective nodes are contending for the channel at the same time. In other words, there is a need to adjust the average number of nodes contending for the channel if their contention areas overlap.

In this paper, we analyze the adjustment to be made in the average number of neighboring nodes contending for the channel. The average number of nodes after adjustment can be used to compute the average delay and/or throughput incurred by packets flowing along a multihop path from a given source to a destination.

The rest of the paper is organized as follows. In Section 2, we present the network model. In Section 3, we analyze the adjustment to be made in the average number of nodes contending for the channel. Section 4 contains results and discussion. Finally, the last section is for conclusion.

## 2   Network Model

Let there be an ad hoc network with $n$ nodes uniformly and randomly distributed in a region of area $A$. Each node is equipped with an omnidirectional antenna and has a transmission radius $r$. Let the signal power at the transmitting node, $P(x = 0)$, be $P_0$. The signal power at a distance $x$ from the transmitting node is given by the following expression.

$$P(x) = \frac{P_0}{x^\gamma}. \tag{1}$$

where, $\gamma$ is called *path loss exponent*, and varies from 2 to 5 depending upon the environment. The transmission range is the maximum distance from the transmitting node up to which the signal can be decoded successfully by the nodes lying in the area of the circle drawn taking the transmitting node as the center of the circle. In other words, it is the maximum distance beyond which the transmitted signal cannot be decoded successfully by a receiver. The signal power at a distance $r = r_0$ is given by

$$P(r_0) = \frac{P_0}{r_0^\gamma}. \tag{2}$$

Beyond the transmission range, the signal power is not strong enough to be successfully decoded by a receiver, however, the signal may collide with signals transmitted by other nodes. The maximum distance from the transmitting node up to which the carrier can be sensed, irrespective of whether it can be decoded or not, is called the *carrier sense range*. The carrier sense range of each node is denoted by $r_{cs}$ and is assumed to be $\beta$ times the transmission range. In other words,

$$\beta = \frac{r_{cs}}{r_0} \qquad (3)$$

where $\beta > 1$, and is generally taken to be equal to 2. Let us call the number of nodes per unit area as the node density of the network i.e. the node density is, $\zeta = \frac{n}{A}$. The number of nodes that are in the transmission range of a node including itself is $\zeta \pi r^2$, and the number of nodes lying in the carrier sensing range is

$$\nu_{cs} = \zeta \pi r_{cs}^2. \qquad (4)$$

The number of neighbors of a node that are lying within its transmission range is $\nu - 1$. Similarly, the number of neighboring nodes lying within the carrier sense range of the node is $\nu_{cs} - 1$. However, the number of nodes contending for the channel including the node itself is $\nu_{cs}$.

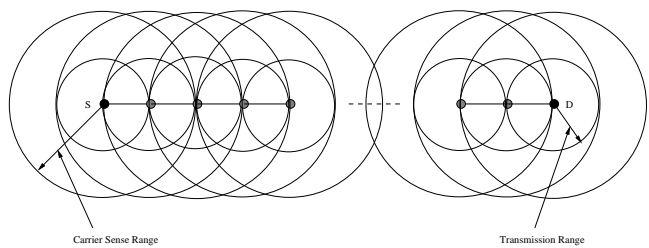

**Fig. 1.** A multihop path between a given source $S$ and the destination $D$. The carrier sensing range of each node is twice of the transmission range.

## 3 Adjustment for The Number of Contending Nodes

In this section, we describe the adjustment to be made for the minimum number of neighbors that are common between the carrier sense ranges of the nodes that are lying along a path from a given source to the destination.

Let there be $h$ hops along a path from a given source to a destination (see Figure 1). It means that there are $h + 1$ nodes forming the path including the source and the destination. However, since the destination does not transmit (for the time instance while it is acting as the destination, and does not have its own frames to transmit), therefore, there are $h$ nodes involved in transmitting the frame.

Figure 1 shows a multihop path from a given source $S$ to a given destination $D$. Source $S$ has a contention with all nodes lying in its carrier sense range. When the frame is in the hands of the first intermediate node along a path, it has a contention with nodes lying in its carrier sense range excluding the nodes lying in the area that is common among the contention areas of the two nodes. This is based on the assumption that queueing delays at the network layer are negligible as compared to the MAC delays, and it seems to be justified for relatively low values of queue load factor[1]. The first intermediate node transmits the frame to the second intermediate node. When the frame is in the hands of second intermediate node, it has a contention with the number of neighbors in its carrier sense range excluding the nodes lying in the area that is common between it and the source. The third intermediate node has a contention with the nodes lying in its carrier sense range excluding the nodes lying in the area that is common between it and the source. The fourth intermediate node has a contention with all nodes lying in its carrier sense range, the fifth intermediate node has contention with nodes lying in the area that is common between it and the fourth intermediate node, and so on.

Assume that we follow the *maximum forward region (MFR)* policy for traversing between successive hops. In other words, there is the maximum advancement while traversing each hop along a path. Further, the paths are assumed to be approximated by straight lines between corresponding source-destination pairs.

We now describe the minimum area that has to be common between two nodes depending upon the distances between their centers.

### 3.1 Common Area Between Two Nodes

Let $t$ be the distance between the centers of two nodes. For $S$ and $i_1$, $t = r$; for $S$ and $i_2$, $t = 2r$; for $S$ and $i_3$, $t = 3r$. For $i_4$ and $i_5$, $t = r$; for $i_4$ and $i_6$, $t = 2r$; for $i_4$ and $i_7$, $t = 3r$; and so on.

An expression for the area of the region which is common between two circles of radii $r$ and $R$, respectively, and whose centres are separated by a distance $t$ is given in [1]. Based on that, the expression for the area that is common between two circles of the same radii $R$, and whose centers are separated by a distance $t$ is as follows.

$$A(t, R) = 2R^2 \left\{ \arccos\left(\frac{t}{2R}\right) - \frac{t}{4R^2} \left(4R^2 - t^2\right)^{\frac{1}{2}} \right\} \tag{5}$$

---

[1] For relatively large values of load factor, the timings of the transmissions of two neighboring nodes might be significantly different due to relatively large queueing delays. In that case, there does not seem to be a need of adjusting the number of neighbors contending for the channel, because each node will, then, contend with all neighbors lying in its carrier sense range. As a result, the end-to-end delay can be taken as the the single hop delay multiplied by the number of hops along the path.

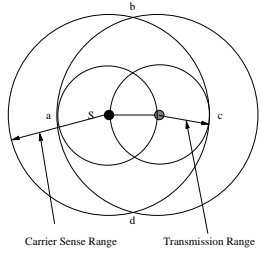

**Fig. 2.** Common area between the carrier sense ranges of two adjacent nodes with a distance between their centers $t = r$.

For $S$ and the first intermediate node, $t = \frac{r_{cs}}{2}$, $R = r_{cs}$, we have,

$$\begin{aligned}
A_{S,i_1} &= A\left(\frac{r_{cs}}{2}, r_{cs}\right) \\
&= 2r_{cs}^2 \left\{ \arccos\left(\frac{r_{cs}}{4r_{cs}}\right) - \frac{r_{cs}}{8r_{cs}^2}\left(4r_{cs}^2 - \frac{r_{cs}^2}{4}\right)^{\frac{1}{2}} \right\} \\
&= 2r_{cs}^2 \left\{ \arccos\left(\frac{1}{4}\right) - \frac{\sqrt{15}}{16} \right\} \\
&= 2.1521 r_{cs}^2 \\
&\approx 2r_{cs}^2 \left\{ \frac{13\pi}{31} - \frac{\sqrt{15}}{16} \right\}.
\end{aligned} \qquad (6)$$

Equation (6) gives the minimum area that is common between two consecutive nodes along a path (say $\mathcal{P}$). Let us call the area $A_{S,i_1}$ as $a_1$. In other words, $a_1$ is the area that is common between node $i$ and $j$, where $j = i+1$ and $i, j \in \mathcal{P}$.

For $S$ and the second intermediate node, $t = r_{cs}$, $R = r_{cs}$, we have,

$$\begin{aligned}
A_{S,i_2} &= A(r_{cs}, r_{cs}) \\
&= 2r_{cs}^2 \left\{ \arccos\left(\frac{r_{cs}}{2r_{cs}}\right) - \frac{r_{cs}}{4r_{cs}^2}\left(4r_{cs}^2 - r_{cs}^2\right)^{\frac{1}{2}} \right\} \\
&= 2r_{cs}^2 \left\{ \arccos\left(\frac{1}{2}\right) - \frac{\sqrt{3}}{4} \right\} \\
&= 2r_{cs}^2 \left\{ \frac{\pi}{3} - \frac{\sqrt{3}}{4} \right\} \\
&= 1.2283 r_{cs}^2.
\end{aligned} \qquad (7)$$

Equation (7) gives the minimum area that is common between two nodes $i$ and $j$ such that $j = i+2$ and $i, j \in \mathcal{P}$. Let us call the area $A_{S,i_2}$ as $a_2$.

For $S$ and the third intermediate node, $t = \frac{3r_{cs}}{2}$, $R = r_{cs}$, we have,

$$\begin{aligned}
A_{S,i_3} &= A\left(\frac{3r_{cs}}{2}, r_{cs}\right) \\
&= 2r_{cs}^2 \left\{\arccos\left(\frac{3r_{cs}}{4r_{cs}}\right) - \frac{3r_{cs}}{8r_{cs}^2}\left(4r_{cs}^2 - \frac{9r_{cs}^2}{4}\right)^{\frac{1}{2}}\right\} \\
&= 2r_{cs}^2 \left\{\arccos\left(\frac{3}{4}\right) - \frac{3\sqrt{7}}{16}\right\} \\
&= 0.4533 r_{cs}^2 \\
&\approx 2r_{cs}^2 \left\{\frac{23\pi}{100} - \frac{3\sqrt{7}}{16}\right\}.
\end{aligned} \tag{8}$$

Equation (8) gives the minimum area that is common between two nodes $i$ and $j$ such that $j = i + 3$ and $i, j \in \mathcal{P}$. Let us call the area $A_{S,i_3}$ as $a_3$.

### 3.2 Common Area Among the Nodes Lying Along a Path

Let the contention area of a node $i$ be denoted by $c_i$. The contention area of a single node is $a = \pi r_{cs}^2$. The contention area that is common among a set of $h$ nodes along a path is given by the following expression.

$$\bigcup_{l=1}^{h} c_l = \sum_{i=1}^{i=h} c_i - \sum_{i=1, j=1, i \neq j}^{i=h, j=h} c_i \cap c_j + \sum_{i=1, j=1, k=1, i \neq j \neq k}^{i=h, j=h, k=h} c_i \cap c_j \cap c_k - \cdots + \cdots$$
$$+ (-1)^{s+1} \sum_{i=1, j=1, k=1, \cdots, m=1, i \neq j \neq k \neq \cdots \neq m}^{i=h, j=h, k=h, \cdots, m=h} c_i \cap c_j \cap c_k \cap \cdots \cap c_m$$
$$+ \cdots + (-1)^{h+1} c_1 \cap c_2 \cap \cdots \cap c_h \tag{9}$$

Or,

$$\bigcup_{l=1}^{h} c_l = \binom{h}{1} c_l - \binom{h}{2} \bigcap_{i \neq j} (c_i, c_j) + \binom{h}{3} \bigcap_{i=1, j=1, k=1, i \neq j \neq k}^{i=h, j=h, k=h} (c_i, c_j, c_k) + \cdots - \cdots$$
$$+ (-1)^{s+1} \binom{h}{s} \bigcap_{i \neq j \neq k \neq m} (c_i, c_j, c_k, c_m) - \cdots + \cdots$$
$$+ (-1)^{h+1} \binom{h}{h-1} \bigcap_{l=1}^{h} c_l. \tag{10}$$

### 3.3 Adjustment Factor

We define a factor, $\chi$, that we call *adjustment factor*, as follows.

$$\nu'_{cs} = \nu_{cs}(1-\chi) \tag{11}$$

where, $\nu'_{cs}$ is the adjusted number of nodes lying in the carrier sense range of a tagged node, and $\nu_{cs}$ is the number of nodes without any adjustment that are lying in the carrier sense range of the tagged node. From (11), we have,

$$\begin{aligned}\chi &= 1 - \frac{\nu'_{cs}}{\nu_{cs}} \\ &= 1 - \frac{\zeta a'}{\zeta \pi r_{cs}^2} \\ &= 1 - \frac{a'}{\pi r_{cs}^2},\end{aligned} \tag{12}$$

where, $a'$ is the area that is common between two or more nodes lying along a path from a given source to a destination.

We now discuss how to evaluate the adjustment factor when there are two nodes, three nodes, ..., six nodes along a path. Let there be two nodes along a path. Using (10), we have

$$\begin{aligned}c_1 \cup c_2 &= c_1 + c_2 - c_1 \cap c_2 \\ &= a + a - a_1 \\ &= 2a - a_1.\end{aligned} \tag{13}$$

The adjustment factor for *two* nodes is

$$\begin{aligned}\chi &= \frac{a_1}{2a} \\ &= \frac{2.1521 r_{cs}^2}{2\pi r_{cs}^2} \\ &= \frac{2.1521}{2\pi} \\ &= \frac{2.1521}{\pi}\left(1 - \frac{1}{2}\right).\end{aligned} \tag{14}$$

Let there be *three* nodes along a path. Using (10), we have

$$\begin{aligned}c_1 \cup c_2 \cup c_3 &= c_1 + c_2 + c_3 - c_1 \cap c_2 - c_2 \cap c_3 - c_1 \cap c_3 + c_1 \cap c_2 \cap c_3 \\ &= a + a + a - a_1 - a_1 - a_2 + a_2 \\ &= 3a - 2a_1.\end{aligned} \tag{15}$$

The adjustment factor for three nodes is

$$\begin{aligned}\chi &= \frac{2a_1}{3a} \\ &= \frac{2 \times 2.1521 r_{cs}^2}{3\pi r_{cs}^2} \\ &= \frac{2 \times 2.1521}{3\pi} \\ &= \frac{2.1521}{\pi}\left(1 - \frac{1}{3}\right).\end{aligned} \qquad (16)$$

Let there be *four* nodes along a path. Using (10), we have

$$\begin{aligned}c_1 \cup c_2 \cup c_3 \cup c_4 &= c_1 + c_2 + c_3 + c_4 - c_1 \cap c_2 - c_1 \cap c_3 - c_1 \cap c_4 - c_2 \cap c_3 \\ &\quad - c_2 \cap c_4 - c_3 \cap c_4 + c_1 \cap c_2 \cap c_3 + c_1 \cap c_2 \cap c_4 \\ &\quad + c_2 \cap c_3 \cap c_4 + c_1 \cap c_3 \cap c_4 - c_1 \cap c_2 \cap c_3 \cap c_4 \\ &= a + a + a + a - a_1 - a_2 - a_3 - a_1 - a_2 - a_1 + a_2 + a_3 \\ &\quad + a_2 + a_3 - a_3 \\ &= 4a - 3a_1.\end{aligned} \qquad (17)$$

The adjustment factor for four nodes is

$$\begin{aligned}\chi &= \frac{3a_1}{4a} \\ &= \frac{3 \times 2.1521 r_{cs}^2}{4\pi r_{cs}^2} \\ &= \frac{3 \times 2.1521}{4\pi} \\ &= \frac{2.1521}{\pi}\left(1 - \frac{1}{4}\right).\end{aligned} \qquad (18)$$

The area that is common between two nodes $i$ and $j$, such that $j = i + 4$ and $i, j \in \mathcal{P}$, is 0. In other words, contention areas of two nodes $i$ and $j$, with $j = i + 4$ and $i, j \in \mathcal{P}$, are disjoint or do not have any common region. Therefore, while evaluating an expression of finding the area of intersection between two or more nodes, e.g. $c_1 \cap c_i \cap c_j$, if there are any two nodes $i$ and $j$ in the expression such that $j = i + 4$, then the value of the contention area represented by the expression is 0. This is exemplified in the ensuing discussion.

In all cases of intersections discussed above, we have not encountered a situation where the area of intersection of two or more nodes is 0. We now discuss the following two cases: (i) five nodes along a path, and (ii) six nodes along a path, where we find that the area of intersection between two or more nodes can

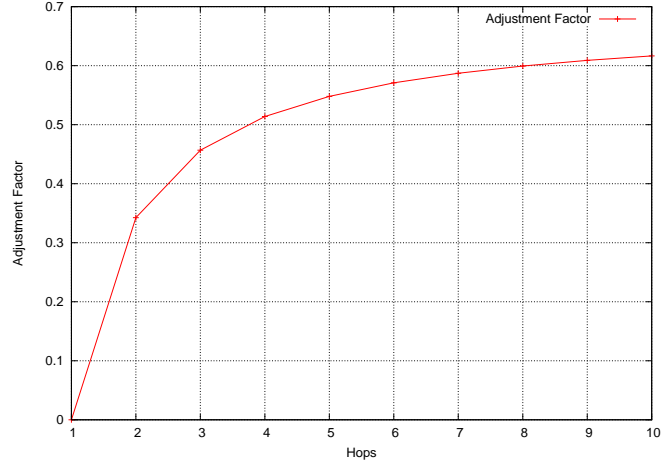

**Fig. 3.** Adjustment factor as a function of the number of hops along a path from a given source to a destination.

be 0. Let there be *five* nodes along a path. Using (10), we have

$$\begin{aligned}
\bigcup_{l=1}^{5} c_l &= c_1 + c_2 + c_3 + c_4 + c_5 - c_1 \cap c_2 - c_1 \cap c_3 - c_1 \cap c_4 - c_1 \cap c_5 - c_2 \cap c_3 \\
&\quad - c_2 \cap c_4 - c_2 \cap c_5 - c_3 \cap c_4 - c_3 \cap c_5 - c_4 \cap c_5 + c_1 \cap c_2 \cap c_3 \\
&\quad + c_1 \cap c_2 \cap c_4 + c_1 \cap c_2 \cap c_5 + c_1 \cap c_3 \cap c_4 + c_1 \cap c_4 \cap c_5 + c_2 \cap c_3 \cap c_4 \\
&\quad + c_2 \cap c_3 \cap c_5 + c_2 \cap c_4 \cap c_5 + c_3 \cap c_4 \cap c_5 - c_1 \cap c_2 \cap c_3 \cap c_4 \\
&\quad - c_1 \cap c_2 \cap c_3 \cap c_5 - c_1 \cap c_2 \cap c_4 \cap c_5 - c_1 \cap c_3 \cap c_4 \cap c_5 \\
&\quad - c_2 \cap c_3 \cap c_4 \cap c_5 + c_1 \cap c_2 \cap c_3 \cap c_4 \cap c_5 \\
&= a + a + a + a + a - a_1 - a_2 - a_3 + 0 - a_1 - a_2 - a_3 - a_1 - a_2 - a_1 \\
&\quad + a_2 + a_3 + 0 + a_3 + 0 + a_2 + a_3 + a_3 + a_2 - a_3 - 0 - 0 - 0 - a_3 + 0 \\
&= 5a - 4a_1.
\end{aligned} \quad (19)$$

The adjustment factor for five nodes is

$$\begin{aligned}
\chi &= \frac{4a_1}{5a} \\
&= \frac{4 \times 2.1521 r_{cs}^2}{5 \pi r_{cs}^2} \\
&= \frac{4 \times 2.1521}{5\pi} \\
&= \frac{2.1521}{\pi}\left(1 - \frac{1}{5}\right).
\end{aligned} \quad (20)$$

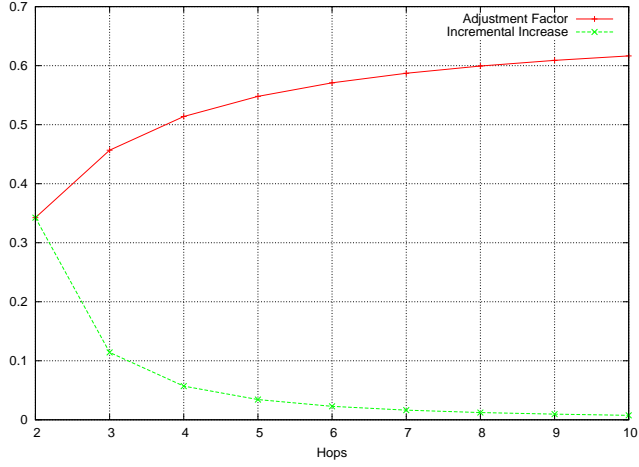

**Fig. 4.** Incremental increase in the adjustment factor as a function of the number of hops along a path from a given source to a destination.

Let there be *six* nodes along a path. Using (10), we have

$$\bigcup_{l=1}^{6} c_l = c_1 + c_2 + c_3 + c_4 + c_5 + c_6 - c_1 \cap c_2 - c_1 \cap c_3 - c_1 \cap c_4 - c_1 \cap c_5$$
$$- c_1 \cap c_6 - c_2 \cap c_3 - c_2 \cap c_4 - c_2 \cap c_5 - c_2 \cap c_6 - c_3 \cap c_4 - c_3 \cap c_5$$
$$- c_3 \cap c_6 - c_4 \cap c_5 - c_4 \cap c_6 - c_5 \cap c_6 + c_1 \cap c_2 \cap c_3 + c_1 \cap c_2 \cap c_4$$
$$+ c_1 \cap c_2 \cap c_5 + c_1 \cap c_2 \cap c_6 + c_1 \cap c_3 \cap c_4 + c_1 \cap c_3 \cap c_5 + c_1 \cap c_3 \cap c_6$$
$$+ c_1 \cap c_4 \cap c_5 + c_1 \cap c_4 \cap c_6 + c_1 \cap c_5 \cap c_6 + c_2 \cap c_3 \cap c_4 + c_2 \cap c_3 \cap c_5$$
$$+ c_2 \cap c_3 \cap c_6 + c_2 \cap c_4 \cap c_5 + c_2 \cap c_4 \cap c_6 + c_3 \cap c_4 \cap c_5 + c_3 \cap c_4 \cap c_6$$
$$+ c_3 \cap c_5 \cap c_6 + c_4 \cap c_5 \cap c_6 - c_1 \cap c_2 \cap c_3 \cap c_4 - c_1 \cap c_2 \cap c_3 \cap c_5$$
$$- c_1 \cap c_2 \cap c_3 \cap c_6 - c_1 \cap c_2 \cap c_4 \cap c_5 - c_1 \cap c_2 \cap c_4 \cap c_6$$
$$- c_1 \cap c_2 \cap c_5 \cap c_6 - c_1 \cap c_3 \cap c_4 \cap c_5 - c_1 \cap c_3 \cap c_4 \cap c_6$$
$$- c_1 \cap c_3 \cap c_5 \cap c_6 - c_2 \cap c_3 \cap c_4 \cap c_5 - c_2 \cap c_3 \cap c_4 \cap c_6$$
$$- c_2 \cap c_3 \cap c_5 \cap c_6 - c_2 \cap c_4 \cap c_5 \cap c_6 - c_3 \cap c_4 \cap c_5 \cap c_6$$
$$+ c_1 \cap c_2 \cap c_3 \cap c_4 \cap c_5 + c_1 \cap c_2 \cap c_3 \cap c_4 \cap c_6$$
$$+ c_1 \cap c_2 \cap c_3 \cap c_5 \cap c_6 + c_1 \cap c_3 \cap c_4 \cap c_5 \cap c_6$$
$$+ c_2 \cap c_3 \cap c_4 \cap c_5 \cap c_6 - c_1 \cap c_2 \cap c_3 \cap c_4 \cap c_5 \cap c_6$$
$$= a + a + a + a + a + a - a_1 - a_2 - a_3 - 0 - 0 - a_1 - a_2 - a_3 - 0 - a_1$$
$$- a_2 - a_3 - a_1 - a_2 - a_1 + a_2 + a_3 + 0 + 0 + a_3 + 0 + 0 + 0 + 0 + 0$$
$$+ a_2 + a_3 + 0 + a_3 + 0 + 0 + a_2 + a_3 + a_3 + a_2 - a_3 - 0 - 0 - 0 - 0$$
$$- 0 - 0 - 0 - 0 - a_3 - 0 - 0 - 0 - a_3 + 0 + 0 + 0 + 0 + 0 + 0 - 0$$
$$= 6a - 5a_1. \tag{21}$$

The adjustment factor for six nodes is

$$\begin{aligned}\chi &= \frac{5a_1}{6a} \\ &= \frac{5 \times 2.1521 r_{cs}{}^2}{6\pi r_{cs}{}^2} \\ &= \frac{5 \times 2.1521}{6\pi} \\ &= \frac{2.1521}{\pi}\left(1 - \frac{1}{6}\right). \end{aligned} \quad (22)$$

**Theorem 1.** *Let $h$ be the number of nodes along a path from a given source to a destination and let the adjusted average number of neighbors lying in the carrier sense range of a node along the path be represented by $\nu'_{cs} = \nu_{cs}(1-\chi)$, where $\chi$ is called the adjustment factor. Then, $\chi$ is given by the following expression.*

$$\chi = \frac{2.1521}{\pi}\left(1 - \frac{1}{h}\right), \quad h \geq 2. \quad (23)$$

*Proof.* We prove it using *principle of mathematical induction*. For $h = 2$, we have,

$$\chi = \frac{2.1521}{\pi}\left(1 - \frac{1}{2}\right), \quad (24)$$

which is true as per (14).

Let the statement of Theorem 1 be true for $h = k$. In other words,

$$\chi = \frac{2.1521}{\pi}\left(1 - \frac{1}{k}\right), \quad k \geq 2. \quad (25)$$

We now add one more node along the path containing $k$ nodes. The area of the union of the set of $k$ nodes is

$$S_k = \bigcup_{l=1}^{k} c_l \quad (26)$$

The area of the union of set of $k$ nodes and the node added to the path is,

$$\begin{aligned}S_k \cup S_1 &= \left(\bigcup_{l=1}^{k} c_l\right) \cup c_{k+1} \\ &= \left(\bigcup_{l=1}^{k} c_l\right) + c_{k+1} - \left(\bigcup_{l=1}^{k} c_l\right) \cap c_{k+1}. \end{aligned} \quad (27)$$

Replacing $\bigcup_{l=1}^{k} c_l$ in (27) by the R.H.S. of (9), and rearranging further, we get an expression for the unions of the contention areas of $k+1$ nodes. Putting the values of contention areas of single, two, three, and four nodes in terms of $a$ and $a_i, i = 1, \cdots, 3$, we get the following expression.

$$\chi = \frac{2.1521}{\pi}\left(1 - \frac{1}{k+1}\right), \quad k \geq 2. \quad (28)$$

Hence, the statement of Theorem 1 is true for all integers $h$.

## 4 Results and Discussion

Figure 3 shows the adjustment factor as a function of the number of hops along a path from a given source to a destination. We observe that as the number of hops increases the adjustment factor initially increases rapidly and then increases slowly. In other words, when the number of hops continue to be larger, the effect of adjustment factor continue to diminish.

Figure 4 shows incremental increase in the adjustment factor as a function of the number of hops along path from a given source to a destination. We observe that as the number of hops are increased, the incremental increase in the adjustment factor decreases. Specifically, when the number of hops in the path are increased from 2 to 3, the incremental increase in the adjustment factor is 33.33%. Similarly, when the number of hops in a path is increased from 3 to 4, the incremental increase in the adjustment factor is 12.5%.

## 5 Conclusion

In this paper, we described a method to adjust the number of neighbors contending for the channel in a multihop wireless ad hoc network that uses a contention-based MAC protocol. We computed the adjustment factor which comes out to be a function of the number of nodes (or hops) along a path from a given source to a destination. We observed that as the number of hops is increased, the incremental increase in the adjustment factor decreases.

## References

1. A.M. Abbas, B.N. Jain, "Topology Control in Heterogeneous Mobile Ad hoc Networks", *Proceedings of 7th IEEE International Conference on Personal Wireless Communications (ICPWC)*, pp. 47-51, January 2005.
2. G. Bianchi, "Performance Analysis of the IEEE 802.11 Distributed Coordination Function", *IEEE Journal on Selected Areas in Communication*, vol. 18, no. 3, pp. 535-547, March 2000.
3. A. Kumar, E. Altman, D. Miorandi, M. Goyal, "New Insights from a Fixed Point Analysis of Single Cell IEEE 802.11 WLANs", *IEEE/ACM Transactions on Networking*, vol. 15, no. 3, pp. 588-601, June 2007.
4. K.R. Duffy, "Mean Field Markov Models for Wireless Local Area Networks", *Markov Processes and Related Fields*, vol. 16, no. 2, pp. 295-328, 2010.
5. I. Tinnirello, G. Bianchi, "Rethinking the IEEE 802.11e EDCA Performance Modeling Methodology", *IEEE/ACM Transactions on Networking*, vol. 18, no. 2, pp. 540-553, April 2010.
6. J. Hui, M. Devetsikiotis, "A Unified Model for the Performance Analysis of IEEE 802.11e EDCA", *IEEE Transactions on Communications*, vol. 53, no. 9, pp. 1498-1510, September 2005.
7. D. Xu, T. Sakurai, H.L. Vu, "An Access Delay Model for IEEE 802.11e EDCA", *IEEE Transactions on Mobile Computing*, vol. 8, no. 2, pp. 261-275, February 2009.
8. K. Kosek-Szott, M. Natkaniec, A.R. Bach, "A Simple But Accurate Throughput Model for IEEE 802.11 EDCA in Saturation and Non-saturation Conditions", *Computer Networks*, vol. 55, no. 3, pp. 622-635, February 2011.